# Understanding the leading indicators of hospital admissions from COVID-19 across successive waves in the UK

Jonathon Mellor[1], Christopher E Overton[1,2,3], Martyn Fyles[1,3], Liam Chawner[1], James Baxter[1], Tarrion Baird[1,4], Thomas Ward[1*]

1. UK Health Security Agency, Data, Analytics and Surveillance, Nobel House, London SW1P 3JR, UK.
2. University of Liverpool, Department of Mathematical Sciences, Peach Street, Liverpool, UK.
3. University of Manchester, Department of Mathematics, Oxford Road, Manchester, UK
4. University of Cambridge, Department of Pathology, Trinity Lane, Cambridge, UK

*Corresponding Author: Tom.Ward@ukhsa.gov.uk

## Abstract

Following the UK Government's Living with COVID-19 Strategy and the end of universal testing, hospital admissions are an increasingly important measure of COVID-19 pandemic pressure. Understanding leading indicators of admissions at National Health Service (NHS) Trust, regional and national geographies help health services plan capacity needs and prepare for ongoing pressures. We explored the spatio-temporal relationships of leading indicators of hospital pressure across successive SARS-CoV-2 incidence waves in England. This analysis includes an evaluation of internet search volumes from Google Trends, NHS triage calls and online queries, the NHS COVID-19 App, lateral flow devices (LFD) and the ZOE App. Data sources were analysed for their feasibility as leading indicators using linear and non-linear methods; Granger causality, cross correlation and dynamic time warping at fine spatial scales. Consistent temporal and spatial relationships were found for some of the leading indicators assessed across resurgent COVID-19 Omicron waves. Google Trends and NHS triages consistently temporally led admissions in the majority of locations, with lead times ranging from 5-20 days, whereas an inconsistent relationship was found for the ZOE app, NHS COVID-19 App, and LFD testing, that diminished with spatial resolution, showing limited cross correlation of leads between -7 to 7 days. This work provides evidence that novel syndromic surveillance data have utility for understanding the expected hospital burden at fine spatial scales. The results indicate that at fine spatial scales novel surveillance sources can be used effectively to understand the expected healthcare burden within hospital administrative areas. However, understanding the temporal and spatial heterogeneity of these relationships is a key determinant of their operational public health utility.

## Introduction

The cessation of mass community testing has hampered the ability to understand contemporary estimates of localised growth of COVID-19. Epidemiological prevalence surveillance studies such as the Office of National Statistics (ONS) COVID-19 Infection Survey (CIS) are produced with inconsistent spatial sampling and significant reporting lags for real-time public health policy. Therefore, the surveillance of hospital admissions and bed occupancy may more reliably capture the true growth of COVID-19 in the community in addition to the current pressures on the health services. Understanding leading indicators of hospital admissions, with improved spatial precision, allows the National Health Service (NHS) to appropriately prepare, and policy makers to plan interventions. Hospital admissions are less influenced by testing ascertainment rates than community testing, however, they are more impacted by the age composition of incidence due to the age severity gradient and are therefore less able to detect early growth in younger ages.

A variety of model structures have been employed to predict COVID-19 admission dynamics. For example, incorporating local testing has been shown to improve admission forecasting at fine spatial scales over autoregressive models [1]. More complex Bayesian structural time series modelling techniques have been developed across several countries to forecast dynamics nationally [2], however, these univariate time series models struggle at epidemic turning points. In addition, causal approaches aimed at capturing herd immunity effects have been employed [5], though these methods rely on assumption-driven scenarios. Transmission modelling has been used throughout the pandemic; however, these models are fit to coarse spatial scales to reduce model uncertainty [6], which limits their operational utility.

There has been work exploring indirect surveillance approaches for disease incidence, for example, perturbations in Google Trend's search terms have been shown to precede cases and deaths at a national level [3] and incorporated in neural network architectures to forecast clinical risk in the UK [4]. These Google Trend search queries also work with other epidemic metrics of interest by improving model predictive performance of case rates, hotspot detection [7], and deaths at a US state level [8]. However, the surveillance data sources are not limited to search engine records; during the COVID-19 pandemic mobility measurements, social media and wearable technology have been explored to forecast healthcare pressures [9 - 12]. These applications are not limited to COVID-19 – digitised syndromic surveillance (including search engines, news reports, social media, clinician search queries, and crowdsourcing apps) have been used effectively to monitor other disease pressures including influenza and zika [13 - 17].

Indicator-admissions temporal relationships are not consistent due to changes in behaviour, population immunological response, testing coverage, and antigenic drift/shift. Novel variants of COVID-19 have had distinct epidemiological characteristics effecting this temporal relationship. The extent of a variant's immunological evasion to prior immunity impacts the incidence growth rate and the rate of spatial dispersion [18]. Novel variants have unique severity profiles [19], for instance relative to wild type, the Alpha variant was estimated to have a 62% (HR – 1.62 (95% CI: 1.48, 1.78)) increased risk of hospitalisation [20]. This evolving relationship between infection and hospitalisation impacts the temporal relationship between indicators and admissions. The COVID-19 vaccination campaign began in the UK in December 2020 reaching 150 million total doses across the first, second, spring and autumn booster doses [21, 22]. The vaccination, in combination with high population infection attack rates over successive waves of SARS-CoV-2 incidence, have led to an increasingly complex picture of immunity at an individual and population level against SARS-CoV-2 infection [23, 24], which impacts syndromic surveillance efforts to understand the spatio-temporal infection burden.

We have evaluated leading indicators of COVID-19 hospitalisations during the Omicron BA.1, BA.2 and BA.4/5 variant waves of 2021/2022. This analysis has been conducted at National Health Service (NHS) Trust geographic scale (local groups of secondary care providers) [25]. We use a variety of methods to assess temporal relationships between indicator and COVID-19 hospital admission at high spatial resolution, including Granger causality, cross correlation analysis and dynamic time warping.

## Methodology

The data assessed were available at different geographic designations and with varying quality. Hospital admission counts by date are provided by the NHS England (NHSE) daily COVID-19 hospital situational report [26]. This contains Trust-level hospital admissions stratified by age, with bed occupancy and staff absence counts. Google Trends [27] was curated to capture search query trends relevant to syndromic surveillance of COVID-19. NHS 111 calls and online pathways [28] were provided by the NHS, with COVID-19 relevant treatment pathways extracted. ZOE Health [29] provided counts of crowd sourced self-reported symptoms from the ZOE App. Lateral Flow Device (LFD) testing data were accessed from the UK COVID-19 dashboard [30]. Aggregated NHS COVID-19 App [31] metrics were extracted from data made available by UKHSA.

Several other data sources were explored for feasibility but were excluded from this study as they were determined to be of limited utility in an operational context. Data sources with a transfer/access latency of greater than one week were excluded, as were data sources that lagged admissions. Primary Care, General Practitioner calls were excluded due to incompleteness of their national spatial coverage. School attendance reports were not evaluated due to the substantial reporting lag and data availability, which hampers utility in operational settings. PCR testing data were excluded due to changes in mass-testing policies which impacted eligibility and spatial coverage. Care home data were excluded due to highly heterogeneous spatial coverage, and a lagging relationship with community transmission. The Office for National Statistics Covid Infection Survey of community positivity was explored and we estimated a high correlation with hospital admission, see *Supplementary Section A, Figure 1*. However, further analysis of this data was not conducted as the quantity of sampled tests in this infection study were not informative at smaller geographic levels (like NHS Trust), nor is it released in a timeframe useful for real-time analysis. Data availability is given in *Supplementary Table 1*. We analysed the data sources in real time, however, for lateral flow device data the analysis was conducted using specimen date which is impacted by data correction over time.

**Data**

**Admissions**

English hospital admissions, provided by NHS England [29], are reported at a Trust level, a collection of secondary care providers. A COVID-19 admission is defined as a patient that had a positive test upon arrival to hospital, or within the past 24 hours while an inpatient. These counts therefore include admissions for COVID-19, incidental presentations and hospital acquired infection. NHSE data are provided daily with each individual Trust submitting the web form by 11am for the preceding 24 hours. Due to the fast operational turnaround and considerable number of hospitals, there are some missing entries per day and occasional inaccurate values reported. The admissions data were reconstructed to provide one record per Trust per day, with resulting missing data being imputed using the last observation carried forward. Organisational mergers were coded manually ensuring records were accurate to the end of the study date. Hospitals with fewer than 10 admissions in 2022 or clear non-acute specialisations were removed from the analysis as they represented either misreporting, non-COVID specialisations or purely incidental admissions, leaving 121 acute Trusts. Admissions are presented from 01 October 2021 to the 29 August 2022, covering the end of the Delta plateau and the BA.1, BA.2 and BA.4/5 Omicron waves in 2021/2022.

**Surveillance Data**

*Google Trends*

A large set of potentially predictive Google Trend's search terms were analysed in previous research to determine which terms should be evaluated [4]. Initially, over 1,000 terms were captured from the most common phrases used within NHS Pathway 111 telephonic COVID-19 triages, COVID-19 symptoms, over the counter medicine and natural language variations on requests for tests. These terms were screened for relevance and relative occurrence at a national level within the Google Trends web interface. Analysis at a national level using generalized additive models with a negative binomial error structure and dynamic time warping were used to assess each term's relevance for COVID-19 incidence.

From the wide range of potential terms, 84 COVID-19 relevant Google Trends search term volume scores were collected hourly and aggregated to daily scores per Lower Tier Local Authorities (LTLA) across the UK England [4]. An LTLA is an administrative geography within the UK, capturing a district of local government, smaller than a county. The Google Trend scores relate to a ranked relative search volume in a city. We transform the city level data to LTLA using a coordinate mapping. The LTLA granularity for the London region is limited, as Google defines London as central London, and then outer London areas, reducing precision when mapping to London LTLAs. Due to the considerable number of Google Search Terms collected, similar terms were grouped together with the aim of increasing the signal to noise ratio. COVID-19 symptom search terms were grouped into "common", "rare" and "severe" terms. Terms relating to general symptoms and tests were combined. General COVID-19 terms such as "coronavirus" were grouped, and terms such as "tier system" relating to policies no longer in effect were combined. The terms and processing logic used are outlined in *Supplementary Section B*.

*NHS 111 Pathways*

NHS 111 provides non-emergency health advice based on individual's symptoms [30]. A user follows the triage process, inputting symptoms and receives healthcare advice (treatment), either through the online or telephone service. Data are presented as counts by day for given pathway outcomes, stratified by age, gender and Lower Super Output Area (LSOA), an area more granular than LTLA. COVID-19 treatments are aggregated together into clinical assessment, ambulance or self-care - stratified by age. The ages were binned into three groups to increase the counts per group. These age groups were: 0-19, 20-59, 60 and over years old.

*LFD Tests*

Lateral flow devices (LFDs) are self-administered rapid tests allowing for real-time detection of COVID-19 by the individual. The tests were provided free universally until 01 April 2022, with the data obtained from the UK COVID-19 Dashboard [31]. An aggregation of positive and total test counts was made available daily at LTLA level. There can be latency in this dataset due to upload delays for the specimen date of test, however, we did not have access to the historical real time data and therefore, analysis included the complete backfilled data. From the LFD data a positivity rate for tests was calculated using positive and total test counts. A metric of test counts per capita was calculated using the associated population size of an NHS Trust.

*NHS COVID-19 App*

The NHS COVID-19 App [32] produces aggregated metrics of app events weekly for analysis. These events covered both app-specific metrics, such as downloads, app store ratings, users, as well as contact tracing-relevant processes such as notifications of exposure. The data were provided at LTLA level with a weekly release schedule within UKHSA, with epidemiologically relevant events – contact exposure notifications of users and reported positive tests via the App interface.

*ZOE COVID-19 Study App*

The ZOE app allows users to self-report COVID-19 relevant symptoms daily with the aim of capturing an up-to-date picture of the pandemic [33]. The data are stratified by age, ethnicity, sex, healthcare worker status and LTLA. Due to the sparsity of counts at this high resolution, the counts were aggregated as total symptom counts per LTLA per day. The different symptom counts were combined into categories, "common", "severe", "rare", and "irrelevant". Groupings for all variables are further described in *Supplementary Section C, Table 2*.

**Processing**

*Spatial Mapping*

NHS 111, ZOE App, LFD tests and NHS COVID App leading indicators are reported in LTLA geographies (or geographies that are strict subregions of LTLAs) and therefore cannot be aggregated directly to hospital admissions at a Trust level. A mapping is therefore required to relate LTLA level data sources to NHS Trust admissions. Using the methodology based on the *covid19.nhs.data* R package [33] we produced a more contemporary probabilistic mapping using count data from the SUS APC (Secondary Use Service, All Patient Care) hospitals admitted patient database. The data extracted was from the 6 months preceding the study and contains test confirmed admissions (with at Trust) and discharge locations (the associated LTLA). From these records we calculate a proportion of people from a local area (LTLA) who were admitted to a specified Trust. Using the proportion of people from an LTLA who attend a Trust, we use the residential population size of that LTLA to calculate the weighted population size for a Trust, referred to as population size in this study. The LTLA residential population counts were obtained from the 2019 mid-year population estimates of each LTLA from the Office for National Statistics. The same mapping between LTLA and Trust allows us to convert LTLA indicator data sources to hospital Trust level. The distribution of populations across NHS Trusts are shown in *Supplementary Figure 2*, compared to counties and LTLAs, showing they are of a comparable scale. For the Google Trends data, a mapping was used which combines the London geographies. Leading indicator sources are first aggregated to LTLA level and then transformed with the Trust-LTLA mapping with a weighted sum to determine the effective impact of an indicator on a Trust.

*Scaling and smoothing*

All indicators were smoothed using a Locally Estimated Scatterplot Smoothing (LOESS) method to reduce noise and scaled between 0 and 1 by Trust to allow comparison between hospitals. For dynamic time warping, the variables were z-score normalized.

**Evaluation**

Data sources are evaluated against present admissions and admissions in 14 days. We look at present admissions to understand the use of the indicator as a real-time proxy for admissions. Following discussion with public health practitioners 14 days was determined as a meaningful time window to act upon leading indicator insight. Consultation on the time length included those creating situational awareness products, senior public health leaders and colleagues within the NHS. Leading relationships less than 14 days are still of interest for surveillance.

To understand how the leading relationships have changed over time the data are analysed across the recent Omicron epidemic waves, with each wave being evaluated independently for the Granger causality and cross correlation approaches. The timing defined for these waves are given in *Supplementary Table 3*.

*Granger Causality*

The Granger causality test estimates if a time series (indicator) can linearly forecast another time series (admissions), not whether there is a causal relationship [34]. The test uses lagged time series and a combination of t-tests and f-tests to determine if the indicator meaningfully adds explanatory power for predicting admissions. Two regression models are constructed, one which contains the explanatory time series (indicator), and one without. The comparison between the two models tells us whether the explanatory time series adds useful information in predicting the response. We have two time series $x_t$ and $y_t$, the indicator and admissions respectively at time $t$. We then construct two regression equations, giving $y_t$ explained by firstly lags of $y_t$ and secondly by lags of $y_t$ and lags of $x_t$

$$y_t = \alpha_0 + \sum_{j=1}^{m} \alpha_j y_{t-j} + \sum_{j=1}^{m} \beta_j x_{t-j} + \epsilon_t \quad (1),$$

$$y_t = \alpha_0 + \sum_{j=1}^{m} \alpha_j y_{t-j} + \epsilon_t \quad (2),$$

where $\alpha_j$ and $\beta_j$ are the regression coefficients for lag $j$. The null hypothesis is that

$$H_0: \beta_1 = \beta_2 = \ldots = \beta_m = 0,$$

and an f-test is performed on the two models (1) and (2) to determine the effect of $x$

$$F = \frac{\left(\frac{RSS_1 - RSS_2}{p_2 - p_1}\right)}{\left(\frac{RSS_2}{n - p_2}\right)},$$

where $n$ is the number of data points, $p_1$ and $p_2$ are the number of parameters in (1) and (2), and RSS is the residual sum of squares. We take the maximum as lag $m = 3$, therefore lags 1, 2 and 3 days are used, with larger numbers of lags reducing the power of the tests. Due to the spatial variation in trends and behaviours at hospital level, Granger causality tests were performed per Trust, rather than at higher aggregations. A test is performed between an indicator and current hospital admissions, as well as between the indicator and hospital admissions in 14 days, to test the relationship within a practically useful temporal distance. Using both times we can understand whether an indicator leads admissions and if the indicator leads admissions enough to be useful.

*Cross Correlation*

A linear time delay analysis using cross correlation functions (CCFs) allow us to calculate the cross correlation between indicators and admissions, producing scores over different lead times [35].

Given two times $x_t$ and $y_t$ where $t = 0, 1, 2 \ldots N - 1$ and that $m_x$ and $m_y$ are the respective means, then the cross correlation $R_{xy}$ at delay $d$ is

$$R_{xy} = \frac{\sum_i [(x_t - m_x) \times (y_{t-d} - m_y)]}{\sqrt{\sum_i (x_t - m_x)^2} \sqrt{\sum_i (y_{t-d} - m_y)^2}}.$$

We define an "optimal lead time" as the lead day $d$, fewer than 30 days, with maximum CCF between indicator and admissions. Within 30 was selected to avoid detecting periodic effects in the growth-peak-decline-plateau cycle of admissions.

*Dynamic time warping*

Dynamic time warping (DTW) calculates the non-linear alignment between two sequences of values. The algorithm creates a mapping (warping curve) between the sequences, which we analyse to understand how indicators and admissions relate over time. The algorithm aims to find the minimal path along the warping curve which aligns the two time series, applied using the R *dtw* package [36]. Further detail on how the method works is provided in [37].

We aim to find the optimal warping curve $\phi$ between $x_i$ and $y_j$. For DTW each index in time series $x_i$ must match at least one time index $y_j$ and the subsequent matches from one index to the next must be monotonically increasing. The algorithm finds the best index matching which minimizes the size of $\phi$, the sum of absolute differences in value for each matched pair of indexes.

$$D(x, y) = \min_{\phi} d_\phi(x, y).$$

The warping between time series can be calculated in a multivariate context (multiple pairs of time series) by minimising the joint distance across the indexes of $x$ and $y$, $i$ and $j$ along the column $C$.

$$d(i, j)^2 = \sum_{c=1}^{C} (x_{ic} - x_{jc})^2.$$

To produce sensible results, we place restrictions on the warping curve. Firstly a 35-day window is used (specifically a Sakoechiba window) to avoid unrealistically long lead times. An asymmetric step pattern with a P2 slope constraint was chosen to allow for a leading alignment between the time series. The specifics of these parameters are addressed in further detail within the DTW literature [37, 38]. From the DTW we can understand the alignment and lead/lag relationship between the time series in a non-linear fashion, identifying leads at different points in time and the epidemic phases by analysing which index pairs are matched. Open start and end conditions were chosen to best capture the leading relationships as not all sequence points of the time series are available, which can cause a beginning or end "bunching" effect, distorting calculated lead times [39]. We analyse the alignments by calculating the difference between the index of the indicator with its matched index in the admissions series, which tells us how far the indicator is ahead of the admissions – a lead time. Additionally, as part of the DTW algorithm, we can extract the cumulative warping $\phi$ between the indicator and admissions, which is a measure of how much manipulation is needed (how different the time series are) to map between the sequences – the warping distance. As time series can be different lengths, or have partial matches, we use the normalized distance to compare how well the indicators match admissions – with greater distances corresponding to larger leads.

*Data Operations*

To be a useful indicator a data source must be available in near-real-time to capitalize on leading information detected and make decisions before the associated admissions occur.

Data sources with substantial reporting latency are not viable candidates for operational indicators. Such a "leading indicator" may have led the measurement in question, but if they are not available to analyse in real time, they cannot have an impact on real-time decisions and were therefore not assessed. Additionally, completion lags, or backfilling, was considered as a major limitation for some data sources, as they are unreliable in the near term.

**Results**

The indicators in this study vary in their magnitude and timing relative to the successive Omicron admission waves – examples of this effect are shown in *Figure 1*.

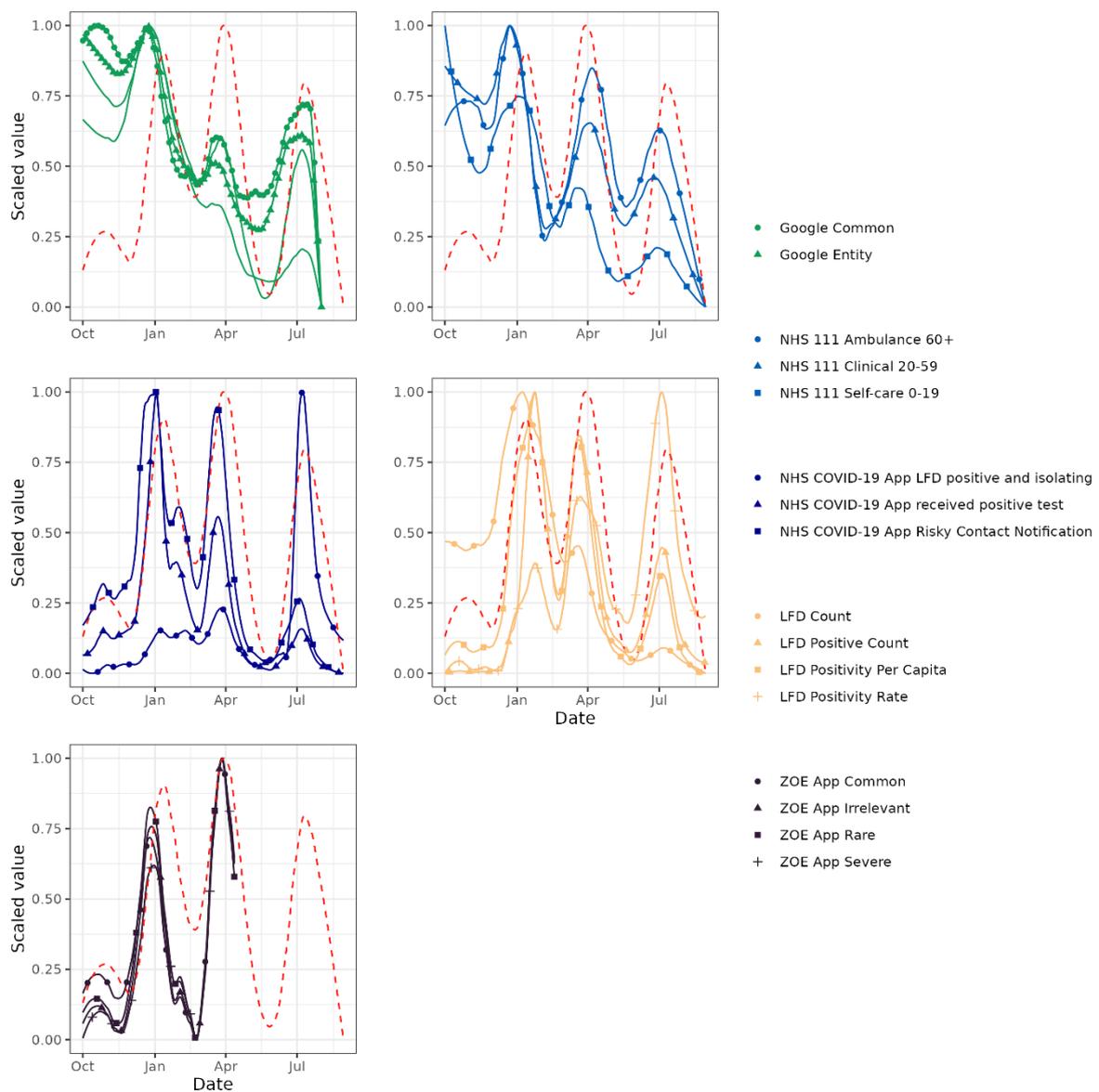

*Figure 1. A time series plot of example variables from each data source's indicators, aggregated nationally, with reference to national admissions (red dashed line). Indicators that lead admissions well should appear shifted leftward of the admissions line. Indicators and admissions are scaled between 0-1 to allow for easy visual comparison of temporal offsets.*

**Granger Causality**

With the assumption of linearity, the Granger tests are used to examine the utility of an indicator in predicting future hospital admissions. A low centred distribution of *p-values* corresponds to an indicator with a strong leading relationship across many Trusts. *Figure 2* describes the *p-value* of the leading relationship of an indicator. The NHS 111 and Google Trends indicators show consistent leading relationships across waves, for some variables. For other indicators such as LFD tests and the NHS COVID App there is high variation in relationship strength across waves. Each wave has unique spatio-temporal indicator relationships, for example the LFD tests were estimated to have the most robust leading association with hospital admissions during the BA.2 period of growth, whereas the strength of association with NHS 111 variables reduced during this wave.

*Figure 3* shows whether an indicator has a lead at 14 days or greater, which is indicative of whether actionable insight can be derived for real-world application. Lower distributed *p-values* indicate a leading relationship at 14 days. Again, there are a majority of Trusts with low *p-values* for the Google and NHS 111 variables, though for the waves available the ZOE symptom data does lead well. However, as the ZOE data are truncated for BA.2, the Granger causality is potentially just capturing the growth phase, which will be easier to linearly predict with Granger causality. Even the strongest indicators have multiple Trusts with high *p-values*, highlighting the spatial variation in relationship.

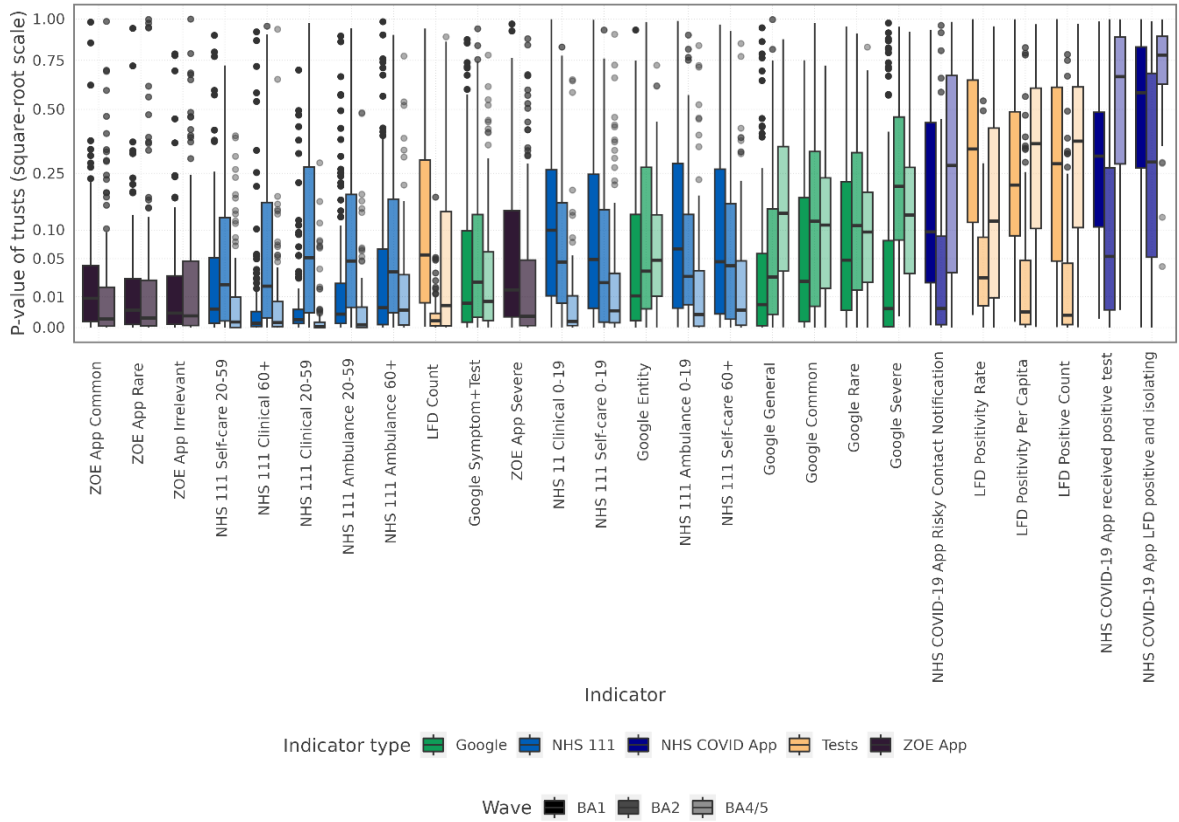

*Figure 2. The distribution of p-values across Trusts from Granger Causality tests. The tests are performed at Trust level for each separate wave between indicator and admissions. Low p-values tell us the indicator leads admissions well given the conditions of the test.*

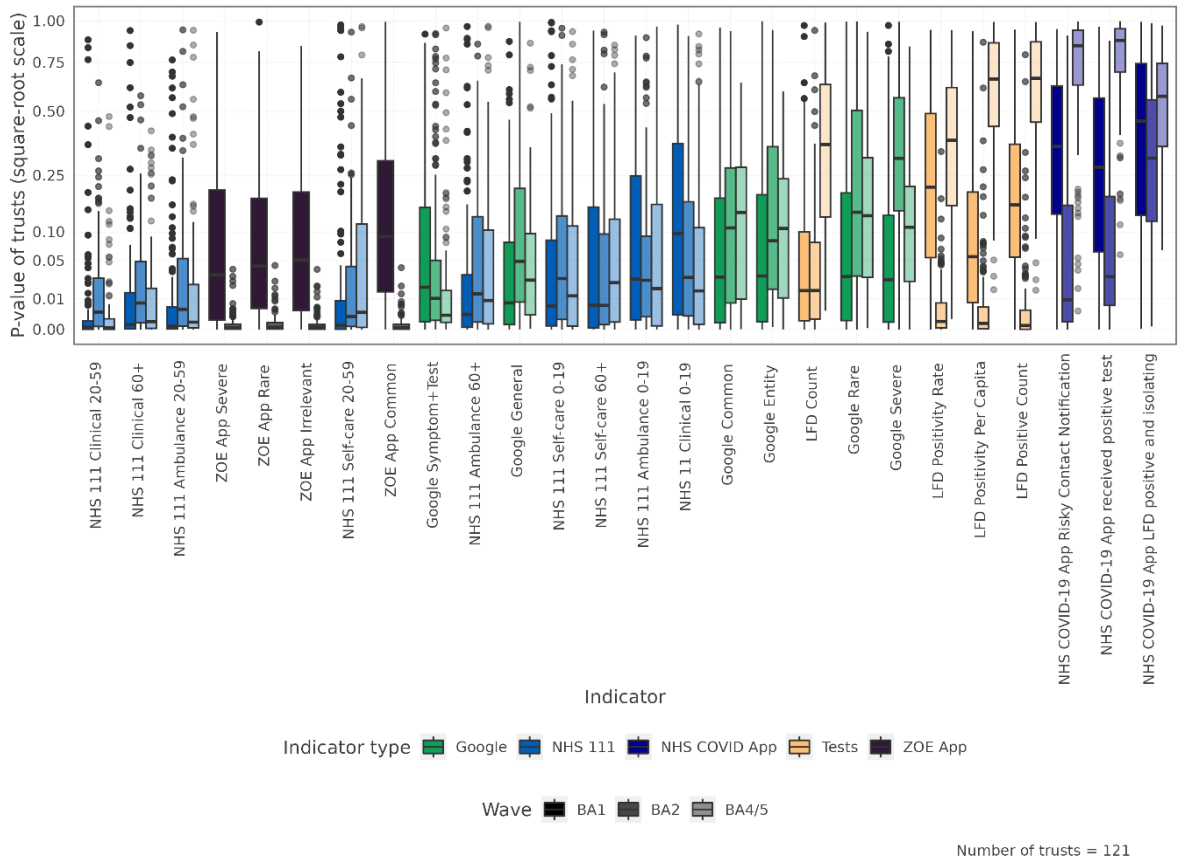

*Figure 3. The distribution of p-values across Trusts from Granger Causality tests. The tests are performed at Trust level for each separate wave between indicator and admissions in 14 days. Low p-values tell us the indicator leads admissions in 14 days well given the conditions of the test.*

**Cross Correlation**

We use cross correlation analysis to assess each indicator at a range of temporal offsets with admissions. We extract optimal lead times and CCF values across Trusts and waves. By calculating the cross correlation between the indicator and admission under linear relationship assumptions at each time offset, we can approximate the optimal lead time, *Figure 4*, and quantify how strong the relationship at 14 days is, *Figure 5*. It is important that we both look at "best lead time" (optimal lead) and the strength at 14 days to understand the nuances of the relationship. Though we are interested operationally in the correlation at 14 days, a given indicator may in fact have a stronger relationship at greater than 14 days, which could be more useful. An indicator may have a large optimal lead time, but that large lead is not useful if the correlation for the indicator is small. Conversely, there may be a small optimal lead, with a highly correlated tracking time series, but high CCF at 14 days, due to a wide peak in CCF values across temporal offsets.

The calculation of optimal lead time is sensitive to its definition, and to perturbation if CCF peaks are wide across lead times. As some proportion of Trusts will not have a leading relationship for an indicator, we expect there to be optimal leads that are either negative (lags) or noisy across waves. An example cross correlation function visualisation is provided in *Supplementary Figure 3*, showing the correlation between NHS 111 Self-care for ages 0-19 and hospital admissions.

An optimal lead time does not necessarily imply a strong leading relationship, it indicates that specific time offset is the lead of highest correlation. For example, as the maximum window is 30 days, maximum correlations at 30 days may indicate a poor signal or out of phase relationship. From an analysis of the distribution of optimal lead times in *Figure 4* we can see there is substantial variation across the waves for individual indicators. For Google Trends and NHS 111 there is a wide range in optimal lead time for the BA.1 wave, whereas for the LFD variables there are instead a lagged relationships, centred below zero. For a given indicator, the sign of the lead can vary substantially across Trusts. While there is a median lead of >10 days for the ZOE symptom variables, the lead reduces in the truncated BA.2 wave, as shown in *Figure 1.*

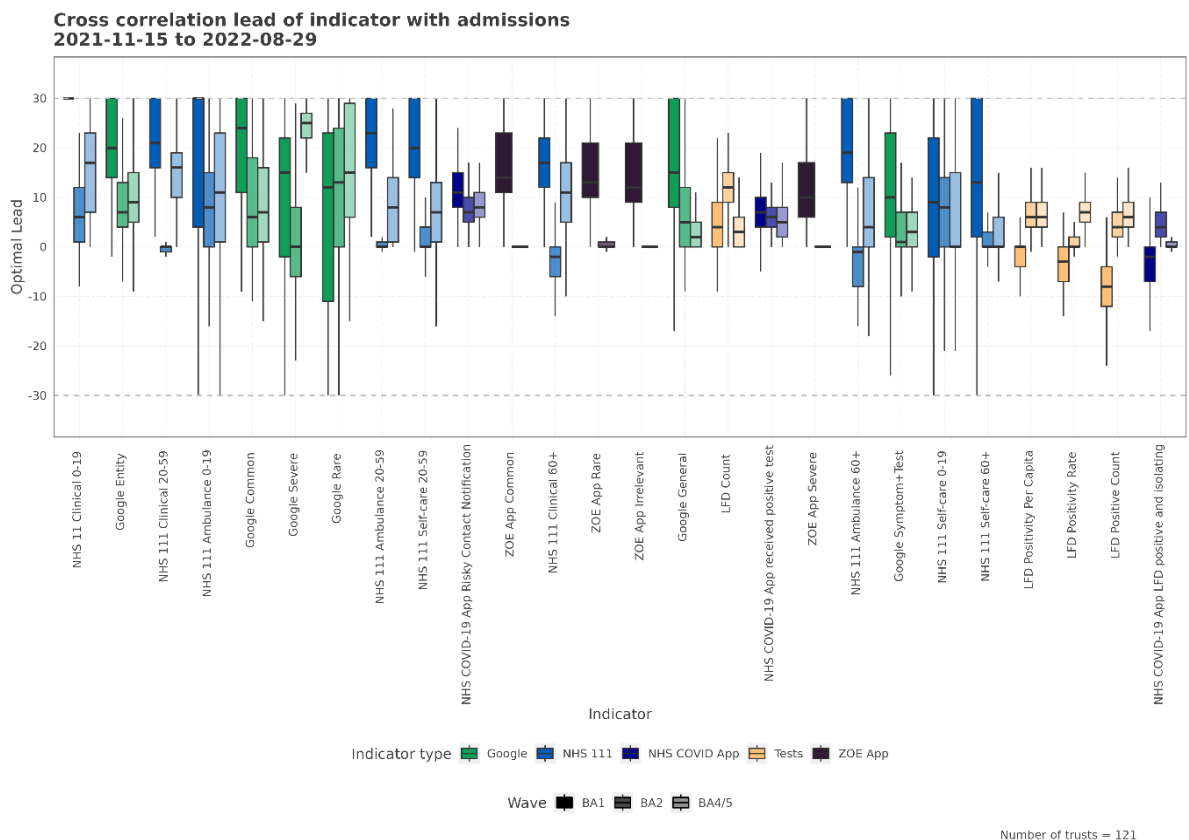

*Figure 4. The distribution, via boxplot, of optimal lead times between the indicator and admissions across all Trusts for each wave. Optimal lead is defined as the lead time with the maximum non-negative CCF value within a 30-day forward and backward window. Larger optimal leads will be most useful for forecasting, wide ranges in optimal lead correspond to high variation spatially, smaller variation indicates a more consistent lead. The indicators with higher average (across waves) optimal leads are sorted to the left.*

By looking at the cross correlation offset at 14 days we can see how related the indicators are with admissions 14 days in the future. A high CCF value indicates a strong relationship at a fixed temporal offset, though variables are likely to have stronger relationships (peak CCFs) at different offsets. We can see the CCF values for the indicators in *Figure 5,* where there is high variation between Trusts and variant waves. Some of the indicators are much more highly correlated than others, though which are the strongest varies wave on wave. Most LFD indicators show stronger relationships in the later waves, for example. There are some negative CCF values, such as NHS 111 Clinical 0-19, which may mean the signal lags admissions - though it may also indicate the signals are out of phase at this specific time offset. The NHS COVID App has consistently strong cross correlation across indicators and waves.

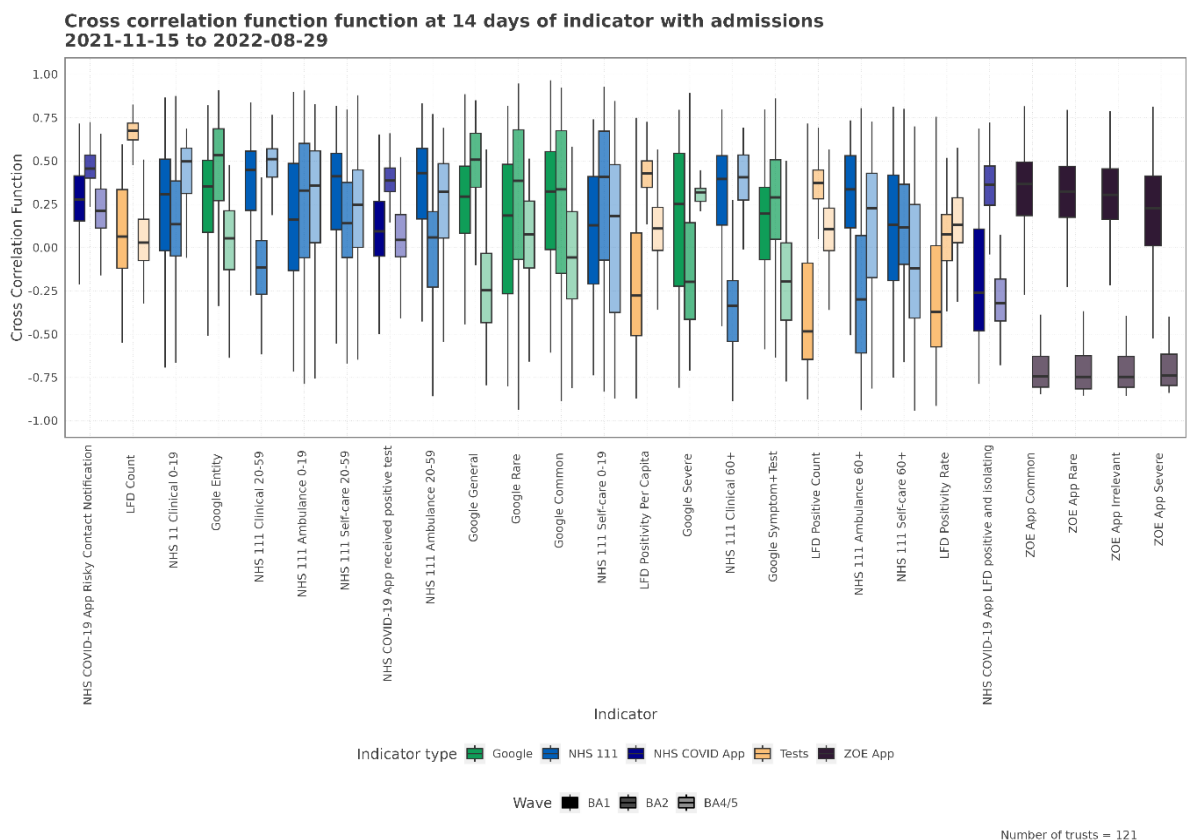

*Figure 5. The distribution of CCF values for the indicator and admissions at fourteen days lead across Trusts for each wave. High CCF values correspond to high correlation between time series, CCF values centred around zero would show that an indicator doesn't have a meaningful temporal lead against admissions. High variation in the CCF values show how consistent leading relationships are across the different Trusts. The indicators with higher average (across wave) CCF values are sorted to the left.*

**Dynamic Time Warping**

We first explore an example dynamic time warping (DTW) for a single Trust with a single indicator to demonstrate the concept, *Figure 6*. There is a clear leading relationship between the indicator and admissions, shown by the diagonal index matches, though the effect decreases in the successive waves. For BA.2 there is a clearer tracking alignment displayed with vertical matches, with peaks at the same indexes for indicator and admission.

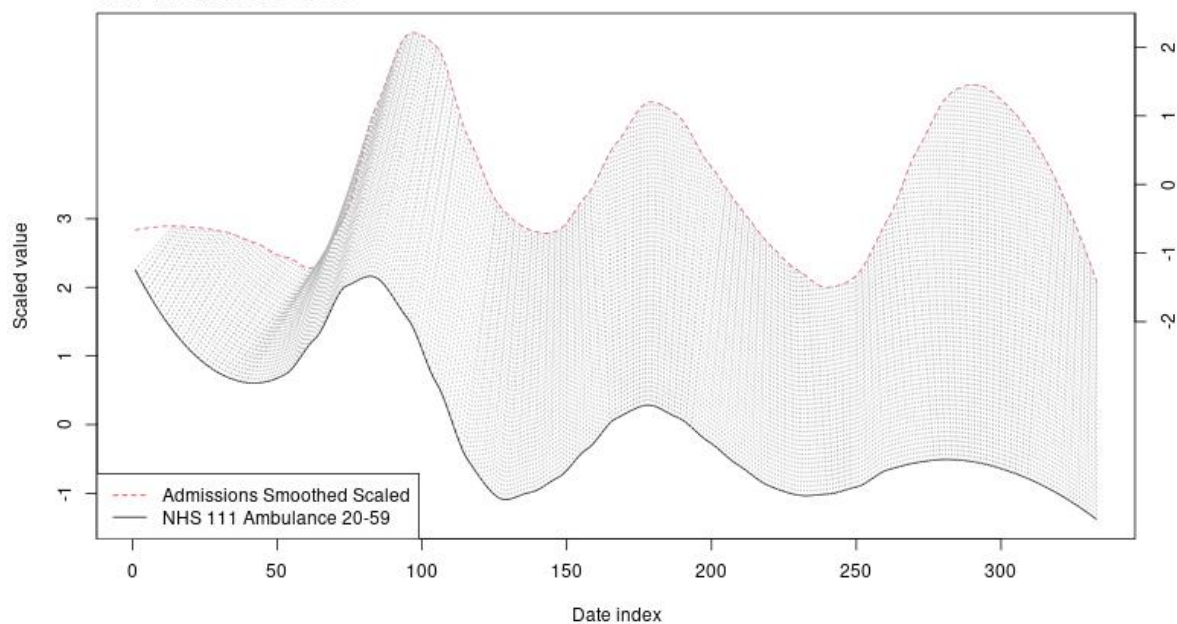

*Figure 6. Dynamic time warping mapping between indicator and admissions used to generate lead times. The DTW is shown for a single indicator and Trust. The solid time series represents the variable being evaluated, the indicator, the dashed are admissions and the lines between are the aligned sequence pairs. Vertical lines indicator no temporal offset between time series.*

For this analysis we focus on the multivariate case of the DTW algorithm, with all Trusts analysed simultaneously for a given indicator, with an example shown in *Supplementary Figure 4.* An initial period of the Delta plateau sequence is included to avoid boundary effects at the start of the series in the DTW algorithm, however, the sequence values are excluded from the reported results. Due to the study time period, an end period following the BA.4/5 wave was not available; therefore, we expect boundary effects to be present in this wave, particularly for the Google Trends, where the time series is further truncated.

In *Figure 7* we show how the alignment lead times are distributed within a wave. Indicators such as Google Trends (Common, Rare, Entity, Symptom + Test), NHS 111 (Clinical 0-19, Ambulance 0-19, 20-59) had leads of greater than 10 days for the BA.1 and BA.2 waves. Using the linear cross correlation approach however, the Google Rare indicator had a smaller lead in BA.2, highlighting how the results are sensitive to approach. For some LFD variables the performance increased over successive waves. For variables with strong leads, the increase was not necessarily replicated for successive waves.

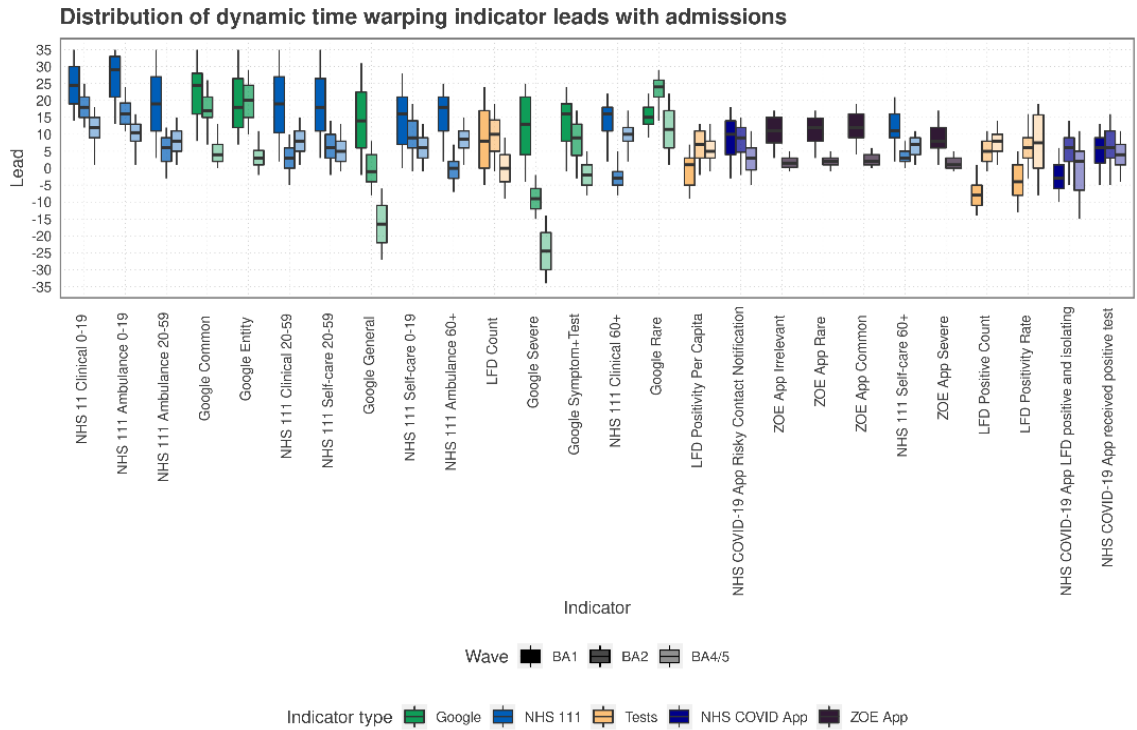

*Figure 7. The distribution of lead times calculated from sequence index matching between indicators and admissions across the different epidemic waves of study. The lead times correspond to the optimal time warping between indicator and admissions, with a higher value indicating a larger temporal lead.*

The normalized distance is a measure of misalignment between the indicator and admissions produced from the dynamic time warping. While the step pattern chosen does not strictly produce a lead alignment for all types of signals, other analysis in this manuscript shows linear lead times > 0, we therefore assume that the normalized distances produced can be treated as a proxy for lead time length. A smaller normalized distance corresponds to a small warp required to align the indicator and admission, which means the indicator tracks the admissions well. Poorly aligned indicators therefore have greater normalized distances and high corresponding lead. Using a normalized distance allows comparison of timeseries of different lengths, such as the ZOE App data and Google Trends, which are truncated. Within *Figure 8* we see the NHS 111 and Google Trends have larger normalized distances compared to other indicators- they are not as aligned. The result may be because they have a large lead/lag, which requires a large warping, or due to having a relationship that is not consistent through time and location. The COVID App, LFD tests and ZOE app have smaller normalized distances indicating a smaller lead/lag or more consistent time shifts across geographies.

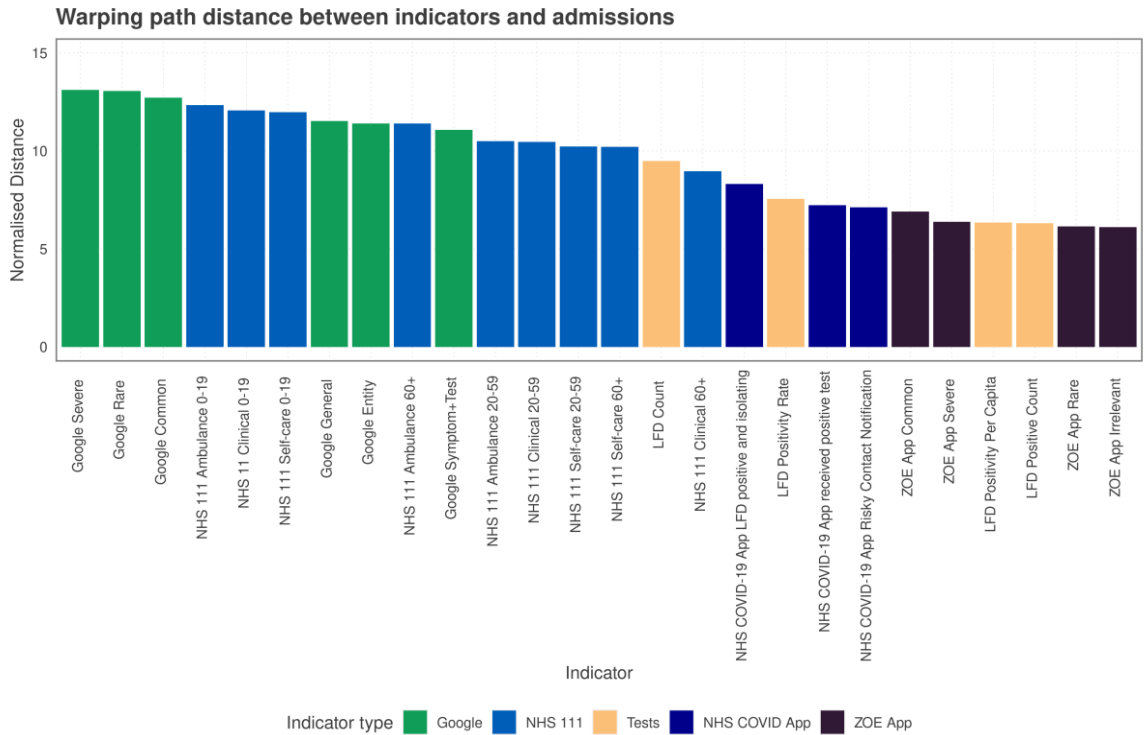

*Figure 8. The normalized path distance produced by warping the space between the indicator and admissions across all three waves. Produced using multivariate DTW across Trusts. The normalized path distance indicates how much total warping is needed between indicator and admissions, a proxy for how big a lead time there is between time series.*

*Data Operations*

To be a useful indicator, data must be available in a timely and complete manner. While leading relationships can be evaluated statistically for historical data, how useful an indicator is in a practical setting is a result of how quickly analysis can be made available. If there are significant reporting/completeness lags in data being presented, then it will erode any actionable lead times. The different release frequencies and reporting/completeness lags are given in *Table 1* for those indicators explored in this study.

| Indicator | Release Frequency | Reporting/completeness lag |
|---|---|---|
| Google Trends | Daily | 1 days |
| NHS 111 Pathways | Daily | 2 days |
| LFD Tests | Weekly | 1 days |
| NHS COVID-19 App | Weekly | 2-3 days |
| ZOE App | Ad hoc | 1 day estimated |

*Table 1. The operational considerations at time of investigation. A 1 day reporting/completeness lag corresponds to yesterdays collected data being available today. More frequent releases allow for more flexibility in times of increased demand for analysis, particularly in a fast-changing epidemic. The reporting/completeness lag "eats into" lead times, reducing how useful a leading indicator is in practice.*

Google Trends is gathered in real time via an API, therefore available without latency, which allows for daily analysis. The NHS 111 data are provided via egress from the NHS, which introduces a day lag. The LFD testing and NHS COVID-19 App's weekly release schedule reduces the operational effectiveness of their use, as an analysis or model may be needed for policy making at different points in the week relative to when the data are released. The one-off provision of ZOE data makes the operational effectiveness and limitations unknown without further exploration. How quickly data can be collected and transferred to analysts directly impacts its utility for decision making.

## Discussion

Using a variety of statistical approaches, we highlight the potential utility of novel surveillance data, displaying their strengths and limitations for use as leading indicators of hospital admissions with COVID-19 at high spatial resolution. The strength of the leading indicator relationship changes temporally and spatially across resurgent waves of COVID-19 incidence and the implications of our results show a range of indicators should be used to accurately capture epidemic trends.

Consistent with the start of the pandemic [11], we show that Google Trends is an effective leading indicator of the healthcare burden associated with SARS-CoV-2 transmission when extended to fine spatial resolutions. Existing work on COVID indicator analysis at fine spatial scales explores cases [1], which we expand upon to show that there are other novel, viable indicators at these scales. Building on the data collection approach used to determine clinical risk at low spatial geographies [4] this analysis shows the Google Trends data's utility can be extended to hospital admissions at a hospital Trust rather than administrative geography. While also analysing leading indicators of hospital admissions, we extend the work of forecasting US hospital admissions in 2021 [40] to show effectiveness of Google Trends as a leading indicator within the Omicron waves.

The ZOE App was created to measure current COVID-19 symptomatic prevalence [32]. We found that there was a temporal lead during the BA.1 wave that degrades by the BA.2 wave – likely due to declining usership. We introduce a novel data source for predicting admissions, the NHS COVID-19 App, which as well as reducing COVID-19 incidence [41] could have utility as a forecasting data source, though its usefulness will depend on continued widespread usage and user notifications [42]. Previous literature shows that univariate epidemic signals have limited reliability across waves [43] which corroborates our findings that there will be differences between waves, admissions, and indicators. This analysis shows that NHS 111 Pathways are a valuable leading indicator of COVID-19 hospital admissions, in agreement with work done within a single NHS Trust [44]. We extend these findings across England, beyond the first COVID-19 wave, and further enrich the data using the 111 Online Pathway.

There is large spatial variation in indicators relationships with Trust level admissions. At higher level geographical scales and in aggregate there are trends, but these trends are not always reliable at a local level – with changing lead times in different waves and locations. The explanation for the change will depend on the indicator (such as testing policy for LFDs, media interest for Google Trends and App use uptake for ZOE and NHS COVID-19 App), which implies the indicators useful in the most recent wave are not guaranteed to work in the next wave – requiring active monitoring. Even within wave there is high variation, as shown by the variation in DTW lead times, which may imply that leading indicators may be more performant at different epidemic phases. As the public conscious and government policies shift away from COVID-19, and other respiratory pathogens resume circulation, we would expect a degradation in the strength of some of these signals.

These findings have implications for developing forecasting models using these indicators. The large spatial variation implies that relying on Trust level indicators alone will perform poorly for a proportion of Trusts. By using pooling techniques via hierarchical modelling, the Trusts with stronger leading indicators could inform those with weaker relationships. The temporal variation shows that we cannot assume a single fixed temporal relationship between indicators and admissions. Therefore, to account for the variation, a range of temporal offsets should be considered by a model for use in prediction. The changing relationship between epidemic waves has multiple implications – primarily that no single indicator signal should be considered entirely temporally and spatially reliable for use in a forecasting model fit to historic data. As the relationships over time change substantially, the effect should either be corrected for, or ad hoc signal exclusions applied. By using a range of indicators, rather than single variables, there is more likely to be a leading relationship in the set of variables chosen. Since individual indicators can be impacted by external events (policy changes, data collection etc.), multiple different sources of indicators should be utilized.

**Limitations and further study**

There are several limitations to this analysis which reflect either data availability or further research required. The Google Trends and ZOE App data were not available, or changed collection format from the start to the end of the period of study, which reduces the reliability of the conclusions drawn from these data and prevents direct comparison with other indicators. The analysis of leading relationships included techniques that assume of linear relationship (Granger-causality and cross correlation). If non-linearity is present, this assumption could provide a partial picture as to whether there is a leading relationship. Further research should employ more complex methods such as kernel approaches to Granger causality [45].

The addition of dynamic time warping to this analysis sheds light on the potential non-linear relationships between indicator and admissions. The dynamic time warping showed a positive lead for the indicators with complete data which supports this study's aim. The scaling and smoothing of the different indicators and admissions will impact the results of the different analysis substantially, since the low-level indicator trends are noisy. The scaling of signals across the whole study period will decrease the accuracy of individual Trust level analysis since the magnitude of the measured indicator changes over time. To further understand the relationships between indicator and admissions the signals should be modelled directly and compared, however, modelling was considered beyond the scope of the study and will be explored in future work. In addition, understanding the factors such as Trust catchment size, acuteness, specialisations, and demographics, which impact the strength of relationship.

Work in this area should focus on how to forecast using spatially and temporally granular data streams for potentially large numbers of indicators. Further utilisation of time series methods would help support the understanding of indicator-outcome relationships and how to best harness them. Substantial amounts of work has looked at leading indicators of healthcare pressures at national or regional levels but understanding how to best use indicators at a finer spatial scale increases the value of these indicators to inform public health systems.

**Conclusion**

This work shows that novel surveillance data sources can be used to reliably understand the expected hospital burden from SARS-CoV-2 transmission. Modelling at a hospital Trust level requires an understanding of the variation across space between the indicators and admissions, as well as the drift of the indicator-admissions temporal leads. We show that clinical, crowd sourced, and healthcare seeking behaviour data at a fine spatial level have positive correlations at lead times, but these relationships are not stable between epidemic waves or constant across England. The different variables selected for analysis were chosen to be as disaggregated as possible whilst maintaining sufficient counts, which allowed us to see indicators at low geographies with strong relationships rather than aggregated signals at higher geographies. This approach was shown to be valuable in understanding the heterogeneity in fine spatial scales. While some data sources and indicators perform better overall our analysis did not identify an indicator that had no substantial variability. Therefore, in practical public health operations spatial and temporal heterogeneity should be considered in a modelling context, avoiding a reliance on single signals which can diverge or degrade from the quantity that is being forecast.

**Conflict of Interest**

The authors have declared that no competing interests exist. The authors were employed by the UKHSA but received no specific funding for this study.

**Data Availability Statement**
UKHSA operates a robust governance process for applying to access protected data that considers:
- the benefits and risks of how the data will be used
- compliance with policy, regulatory and ethical obligations
- data minimisation
- how the confidentiality, integrity, and availability will be maintained
- retention, archival, and disposal requirements
- best practice for protecting data, including the application of 'privacy by design and by default', emerging privacy conserving technologies and contractual controls

Access to protected data is always strictly controlled using legally binding data sharing contracts.

UKHSA welcomes data applications from organisations looking to use protected data for public health purposes.

To request an application pack or discuss a request for UKHSA data you would like to submit, contact DataAccess@ukhsa.gov.uk.